\documentclass[aps,prl,twocolumn,showpacs,citeautoscript,superscriptaddress]{revtex4-1}

\usepackage{graphicx}  
\usepackage{dcolumn}   
\usepackage{bm}        
\usepackage{amssymb}   
\usepackage[none]{hyphenat}  
\usepackage{nicefrac}  
\usepackage{amsmath}   
\usepackage{footmisc}  
\usepackage{latexsym}  
\usepackage{romannum}  
\usepackage{hyperref}  
\usepackage{multirow}  

\renewcommand{\arraystretch}{1.5}

\begin{document}

\title{Direct thermal infrared vision via nanophotonic detector
  design}

\author{Chinmay Khandekar} \email{ckhandek@stanford.edu}
\affiliation{Ginzton Laboratory, Department of Electrical Engineering,
  Stanford University, Stanford, California 94305, USA}

\author{Weiliang Jin} 
\affiliation{Ginzton Laboratory, Department of Electrical Engineering,
  Stanford University, Stanford, California 94305, USA}

\author{Shanhui Fan} \email{shanhui@stanford.edu}
\affiliation{Ginzton Laboratory, Department of Electrical Engineering,
  Stanford University, Stanford, California 94305, USA}

\date{\today}

\begin{abstract} 
  Detection of infrared (IR) photons in a room-temperature IR camera
  is carried out by a two-dimensional array of microbolometer pixels
  which exhibit temperature-sensitive resistivity. When IR light
  coming from the far-field is focused onto this array, microbolometer
  pixels are heated up in proportion to the temperatures of the
  far-field objects. The resulting resistivity change of each pixel is
  measured via on-chip electronic readout circuit followed by analog
  to digital (A/D) conversion, image processing, and presentation of
  the final IR image on a separate information display screen. In this
  work, we introduce a new nanophotonic detector as a minimalist
  alternative to microbolometer such that the final IR image can be
  presented without using the components required for A/D conversion,
  image processing and display. In our design, the detector array is
  illuminated with visible laser light and the reflected light itself
  carries the IR image which can be directly viewed. We realize and
  numerically demonstrate this functionality using a resonant
  waveguide grating structure made of typical materials such as
  silicon carbide, silicon nitride, and silica for which lithography
  techniques are well-developed. We clarify the requirements to tackle
  the issues of fabrication nonuniformities and temperature drifts in
  the detector array. We envision a potential near-eye display device
  for IR vision based on timely use of diffractive optical waveguides
  in augmented reality headsets and tunable visible laser sources. Our
  work indicates a way to achieve direct thermal IR vision for
  suitable use cases with lower cost, smaller form factor, and reduced
  power consumption compared to the existing thermal IR cameras.
\end{abstract}

 \maketitle
  
\section{Introduction}
  
\begin{figure}[t!]
  \centering\includegraphics[width=\linewidth]{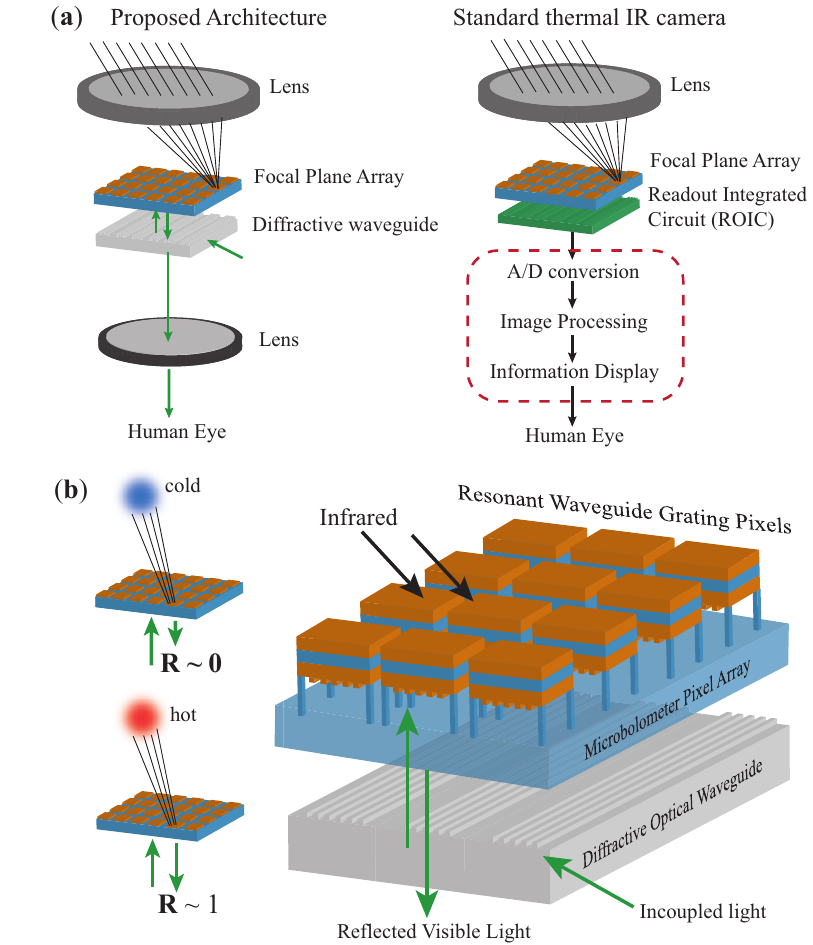}
  \caption{(a) The proposed architecture is schematically shown in
    comparison to the standard thermal IR imaging system. The IR light
    from the far field is focused onto the array of detector pixels
    (also referred to as FPA) which are suspended on top of a
    substrate. The absorption of IR photons causes heating of these
    suspended detector pixels. In the standard design (right panel),
    the detector pixel is a bolometer whose resistivity changes
    sensitively with its temperature variation relative to the
    substrate. The resistivity change is measured using an electronic
    readout circuit (ROIC). After additional A/D conversion and image
    processing, the final IR image is presented on a separate display
    screen. We propose a new design (left panel) where the final IR
    image can be directly visualized via reflection of visible laser
    light illuminating the FPA. We realize this functionality by
    designing the detector in such a way that its reflectivity changes
    from low to high in response to IR-absorption-induced heating as
    illustrated in (b) so that cold and hot objects in the far field
    can be distinguished. In particular, our detector is a resonant
    waveguide grating structure illustrated by the right schematic in
    (b). Its underlying working principle is based on sensing small
    temperature changes of the detector pixel caused by thermo-optic
    effect. Diffractive optical waveguide is used to incouple visible
    light and illuminate the detector array as shown. The wavefront of
    the reflected light spatially modulated by the temperature map of
    the detector pixels carries the final IR image which can be
    directly viewed.}
  \label{schematic}
\end{figure}

Thermal infrared cameras provide images at infrared wavelengths,
typically in the long-wave infrared (LWIR) range of $8\mu m$ to $14\mu
m$, thus allowing us to transcend the visible-spectrum limitation of
human eyes~\cite{gade2014thermal}. The infrared cameras consist of an
imaging system and infrared detectors of which there are two types,
namely photon
detectors~\cite{tan2018emerging,zhuge2017nanostructured,lapierre2017review}
and thermal
detectors~\cite{jayaweera2008uncooled,kruse1997principles}.  The
former rely on the conversion of incident IR photons to electrons by
using low-bandgap semiconductor detectors which typically require low
cryogenic temperatures. The latter rely on the conversion of incident
IR photons to heat using microbolometer detector pixels. The converted
heat results into small temperature variations which can then be
measured to form an image. Such thermal detectors can operate at or
near room temperature. Based on operating temperatures, photon
detectors and thermal detectors are also called as cooled and uncooled
IR detectors respectively. The advantages of thermal detectors over
photon detectors are lower cost, smaller size, lighter weight, lower
maintenance, longer lifetime, immediate power-up capability and
reduced power consumption, while the disadvantages are lower
sensitivity and intrinsically slower response speed (typical time
constant of few $ms$). Despite some of these disadvantages, the
thermal detectors and cameras are used for an increasing number of
cost-sensitive applications that do not necessarily demand high
performance and speeds of photon
detectors~\cite{kimata2018uncooled,takasawa2015uncooled}.
Microbolometers are now produced in larger volumes than all other IR
detector array technologies together~\cite{rogalski2017next}.

In this work, we introduce a new nanophotonic detector in place of
microbolometer to considerably simplify the overall architecture of
the thermal IR imaging system, where `thermal' indicates the
underlying thermal detector as well as our focus on the detection of
LWIR radiation generated thermally in the far-field. Our proposed
design is schematically illustrated in the left panel of
Fig.~\ref{schematic}(a) along with the schematic design of the
standard thermal IR imaging system in the right panel (with dimensions
not to scale). In both cases, LWIR radiation from the far field is
focused by an infrared-transparent lens onto the focal plane array
(FPA) which consists of micron-size detector pixels suspended on top
of a substrate and designed to be good LWIR absorbers. In the standard
design based on bolometers, such a pixel contains a phase change
material (e.g. vanadium oxide) having temperature-sensitive electrical
resistivity. A bias current is passed through the pixel and the
voltage difference across it is measured using a readout integrated
circuit (ROIC). The IR-absorption-induced temperature change of the
bolometer causes its resistivity to change (typically few $\%/K$). The
resulting voltage change signals measured by ROIC are fed to A/D
converters and after additional digital processing, final IR image is
presented on a separate information display screen.

Our design is motivated by the idea of directly visualizing the pixel
temperature changes relative to the substrate by measuring the
reflection of visible light. This will eliminate the need for A/D
conversion, image processing, and a separate display screen (red box).
In this architecture, visible light is projected onto the array of
detector pixels using a diffractive optical waveguide which is used in
augmented reality eye-wearables that overlay digital content on top of
the real-world scene~\cite{kress2021waveguide}. As illustrated in
Fig.~\ref{schematic}(b), the detector pixel is designed such that large
temperature changes relative to the supporting substrate corresponding
to hot objects in the far field lead to high reflectivity ($R \approx
1$) while small temperature changes corresponding to cold objects in
the far field lead to low reflectivity ($R \approx 0$). The light
reflected from the pixel array is spatially modulated by the map of
the temperature changes of the detectors. The diffractive optical
waveguide is usually designed with a small diffraction efficiency,
such that the reflected light from the pixel array can transmit
through the waveguide layer in the perpendicular direction. The
resulting temperature map of $\sim mm^2$ spot size can be directly
viewed through the magnifying lens as shown in
Fig.~\ref{schematic}(a). While many parts of this optical architecture
currently exist, the detector pixel for performing the intended
functionality needs to be specially designed, and that is the focus of
this work.


We note that optical measurement of thermal IR detector pixels has
been explored previously in
Refs.~\cite{ostrower2006optical,pris2012towards,shen2018bioinspired,
  watts2007microphotonic,exner2013low,zhao2018plasmo,choi2003design}.
Moreover, Refs.~\cite{pris2012towards} and ~\cite{choi2003design} also
show the use of charge-coupled device (CCD) imagers to form an image
based on light reflected from the detector pixels. In contrast to
these works, here we envision that the temperature map on the detector
pixel array to be directly viewed by human eyes through a magnifying
lens, in order to further simplify the design of thermal imaging
systems. For this purpose, we design the pixels to achieve low to high
reflectivity change at visible illumination wavelength indicating high
contrast between cold and hot objects in the far field as illustrated
in Fig.~\ref{schematic}(b). Our design utilizes resonant waveguide
gratings~\cite{rosenblatt1997resonant,quaranta2018recent} which are
different in terms of both geometry and material options from all
above-mentioned
works~\cite{ostrower2006optical,pris2012towards,shen2018bioinspired,
  watts2007microphotonic,exner2013low,zhao2018plasmo,choi2003design}
in the context of optical measurement of IR detectors.

The commercial thermal IR cameras, which have matured after many
decades of research~\cite{rogalski2017next}, primarily use bolometers
with electronic ROIC measurements. Although comparable or better
performance may be achieved, introducing optical measurement instead
of electronic readout in this well-established technology is
challenging unless there is a significant (order of magnitude) or a
distinct technological advantage of doing so. Our work advances this
line of research by bringing to attention the unique advantage of
optical readout in enabling direct thermal IR vision with a carefully
engineered system such that additional components for A/D conversion,
image processing, and a separate information display are no longer
required. Since electronic readout is not sufficient for presenting
the final IR image and these additional components are necessarily
required for standard thermal IR cameras, the capability to realize
direct thermal IR vision is in principle a distinct advantage of
optical readout over electronic readout. Our work can potentially help
to further advance the thermal IR imaging technology by reducing the
form factor (size, weight), cost, and power consumption. It can pave
the way for developing small-size, light-weight eye-wearables that can
potentially augment human vision with thermal infrared vision.

\section{Results}

The proposed functionality illustrated in Fig.~\ref{schematic}(b)
requires each detector pixel to satisfy the following
conditions. First, the pixel's reflectivity $R$ at the visible
illumination wavelength should change substantially with its
temperature changes relative to the substrate. Moreover, it should
change from low to high for increasing temperature differentials to
indicate contrast between cold and hot objects in the far
field. Second, the pixel should be a good absorber around the
wavelength of $10\mu$m which corresponds to the peak wavelength of
thermal radiation from near-room-temperature bodies. Third, the pixel
should have good thermal isolation from its surroundings so that the
absorbed thermal radiation can lead to a reasonable change in
temperature, typically of the order of few Kelvins compared to the
substrate.  Fourth, the thermal time constant of the pixel should be
small so that the response speed to the temporal variation of the
incident thermal radiation is fast. Since the third requirement above
limits the thermal conductance of the pixel to the surrounding
environment, the thermal mass of the pixel should be sufficiently
small. For standard thermal IR cameras, the third requirement is
addressed by the suspension of microbolometer pixels on top of a
substrate inside vacuum packaging such that the heat exchange occurs
primarily via conduction between the detector and the
substrate. Additionally, radiation shields are used to prevent any
unwanted IR radiation from the packaging which is unrelated to the
far-field scene~\cite{garcia2017radiometric}. In the following,
assuming the use of vacuum packaging and radiation shields, we
describe our nanophotonic design that addresses the four pixel
requirements. We also analyze the effects of any spurious temperature
drifts and potential fabrication nonuniformity within or across
detector pixels on the device performance, and clarify the related
requirements and calibration methods to tackle them.

\begin{figure}[t!]
  \centering\includegraphics[width=\linewidth]{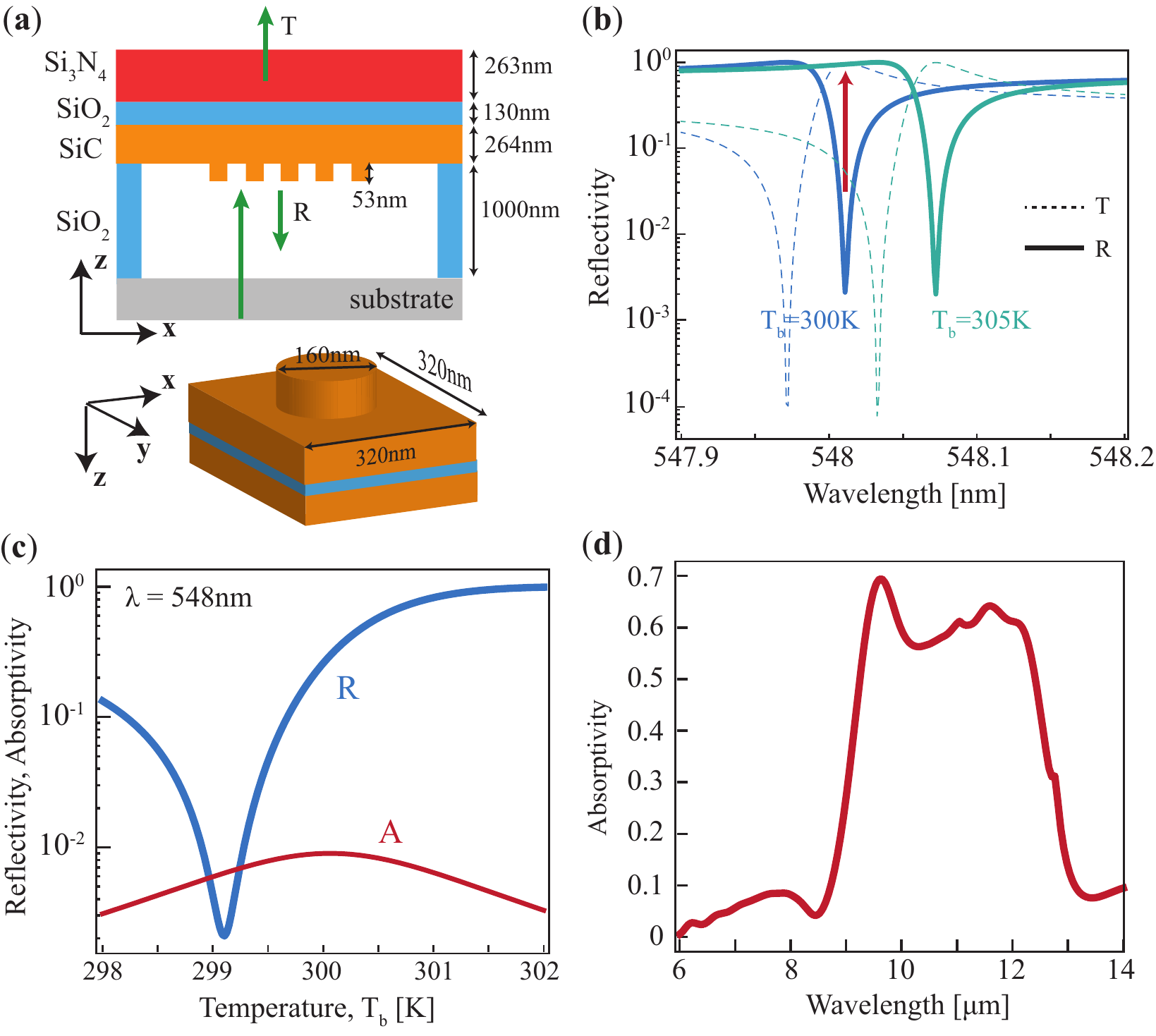}
  \caption{(a) Design of suspended photonic detector pixel whose
    reflectivity is highly sensitive to its temperature change. (b)
    The reflectivity ($R$) and transmission ($T$) spectra at
    $T_b=300K$ and $T_b=305K$. The red arrow indicates the
    illumination wavelength where large reflectivity change can be
    obtained. (c) $R$ as a function of $T_b$ shows strong variation
    over the small temperature change arising from the absorption of
    incident IR light. The absorptivity $A$ is also plotted in the
    same range. (d) The IR absorption spectrum of the detector pixel
    shows broadband nature in the LWIR range of $8-14\mu m$.}
  \label{fig1}
\end{figure}

Figure~\ref{fig1}(a) illustrates the pixel design (dimensions not to
scale). It consists of a multi-layered patterned slab depicted in the
top inset consisting of layers of silicon nitride (Si$_3$N$_4$),
silicon dioxide (SiO$_2$), silicon carbide (SiC) with patterning in
the form of a two-dimensional lattice of cylinders as depicted in the
bottom inset. This constitutes a detector pixel which is suspended on
top of four insulating posts on a substrate. For calculations below,
we assume substrate to be silica glass. The individual layer
thicknesses and the lattice periods are mentioned in the schematic in
Fig.~\ref{fig1}(a). We assume that each pixel is $12\mu m$ $\times
12\mu m$ in lateral directions and separated by $1\mu m$ from the
neighboring pixel. Such pixel dimensions are common in photon IR and
thermal IR detectors with the latest research reporting pixel sizes
down to $5\mu m$~\cite{rogalski2016challenges}. Since refractive
indices of SiC and Si$_3$N$_4$ are larger than that of SiO$_2$ at
visible wavelengths, these layers support guided modes which can
couple to normally incident light via periodic grating, resulting into
sharp resonant features in reflection and transmission
spectra~\cite{wang1993theory,tibuleac1997reflection,fan2002analysis,
  crozier2006air}. Because the materials are low-loss and transparent
in this wavelength range, it is possible to realize high quality
factor ($Q \gtrsim 10^4$) resonances. Such guided mode resonance
structures also identified as resonant waveguide gratings
(RWG)~\cite{rosenblatt1997resonant,quaranta2018recent} and analogous
high contrast gratings with slight geometric
modifications~\cite{chang2012high} have been well-studied and also
used for sensing small changes in refractive
index~\cite{fang2006resonant,xu2019optical}. However, their use for
the optical readout of IR detectors was not previously considered.

Figure~\ref{fig1}(b) demonstrates the reflection (solid lines) and
transmission (dashed lines) spectra for normal incidence of light
around $\lambda = 550$nm and for two different detector temperatures
$T_b=300$K and $T_b=305$K. The small temperature change induced by the
absorbed thermal IR signals from the far-field causes a small change
in the refractive index via thermo-optic effect. The thermo-optic
coefficient of SiC given by $\frac{\partial n_{\text{SiC}}}{\partial
  T} = 5.8\times 10^{-5}/K.$ is much larger than the coefficients of
SiO$_2$ and Si$_3$N$_4$ (of the order of
$10^{-6}/K$)~\cite{watanabe2011thermo,powell2020high}. As shown in
Fig.~\ref{fig1}(b), the high $Q$-factor resonant dip in the reflection
spectrum shifts noticeably toward larger wavelength upon increasing
the pixel temperature. When this structure is illuminated by a laser
source close to $\lambda=548$nm (green light), a large change in the
reflectivity is observed as indicated by the red arrow. $R\approx
10^{-3}$ at $T_b=300K$ changes to $R\approx 1$ at $T_b=305K$. The
temperature dependence of the reflectivity $R$ is demonstrated in
Fig.~\ref{fig1}(c) assuming normal incidence of monochromatic laser
light at $548$nm. The plot indicates that the reflectivity $R$
increases monotonically as the temperature $T_b$ increases from
$299.1K$ up to $302K$. Thus, if the substrate temperature is fixed at
$T_s=299.1K$, this specific example system with illumination at
$548$nm can be used to sense the differentials of $(T_b-T_s)$ induced
by IR absorption. Although we fixed $T_s$ and illumination wavelength
for explaining this underlying mechanism, we explain further below
that the illumination wavelength can be readily tuned in real time by
sensing the substrate temperature such that $R \approx 0$ in the
absence of any incident thermal IR signals. We also note that the
laser illumination can cause the detector pixel to heat up depending
on illumination intensity, pixel area and pixel absorptivity. The
proposed structure exhibits low absorptivity ($A\approx 10^{-2}$) as
shown by the red curve in Fig.~\ref{fig1}(c). For our specific design,
we show below that the heating induced by the laser for typical
illumination intensity appropriate for human vision is much smaller
than the typical absorbed thermal IR power because of this low
absorptivity.


We now discuss the details of our design related to the problem of IR
detection. The thicknesses and lattice periods in the proposed design
are obtained using optimization informed by the physics of RWG
structures as described in the methods section. The example system
discussed above is by no means unique, and many solutions at various
visible wavelengths can be readily obtained for different geometric
parameters (thicknesses and lattice periods) using the same material
layers. We also note that this optimization is performed with respect
to the optical readout and not with respect to thermal IR
absorption. Nonetheless, the proposed pixel structure is already a
reasonably good absorber of thermal radiation as shown in
Fig.~\ref{fig1}(d) which is ensured by the specific choice of
materials. In particular, the material choice in our design is
governed by the fulfillment of two conditions, namely good infrared
photon absorption in $8-14\mu m$ range for LWIR detection and low-loss
transparent nature at visible wavelengths ($\sim 300-700 nm$) for
realizing high $Q$-resonance to detect small changes in pixel
temperature.  In microbolometers used in standard thermal IR
detectors, Si$_3$N$_4$ is already used because of its broadband LWIR
absorption. While SiC and SiO$_2$ also lead to resonant absorption in
$8-12\mu$m due to their optical phonon polaritons, the resulting
absorption spectrum is rather narrowband. Therefore, addition of
Si$_3$N$_4$ layer in the present design helps enhance and broaden the
LWIR absorption. In standard microbolometers, phase-transition
materials such as vanadium oxide (VO$_x$) and amorphous silicon (aSi)
which exhibit large thermal coefficient of resistance (TCR) of few
$\%/K$ are typically used. While the same materials also exhibit large
thermo-optic effect, they are quite lossy at visible
wavelengths. Therefore, these materials as well as metals and metal
oxides cannot be used for the intended detector pixel design.

Below, we quantitatively describe the thermal aspects of our
nanophotonic design. The temperature of the detector pixel follows the
heat equation:
\begin{align}
  m_t \frac{dT_b}{dt} = -G(T_b-T_s)+ A I_b A_p + P_{\text{t}}
  \label{heat}
\end{align}
where $m_t$ is the thermal mass given by the product of the mass and
the specific heat capacity of each material layer, $G$ is the thermal
conductance between the pixel and the substrate via insulating posts,
$I_b$ is the intensity of the laser illumination, $A \lesssim 10^{-2}$
is the absorptivity plotted in Fig.~\ref{fig1}(c), $A_p$ is the pixel
area, $P_{t}$ is the absorbed IR power. Using mass densities
$\rho_{\text{SiC}}=3210 kg/m^3$, $\rho_{\text{SiO$_2$}}=2650 kg/m^3$,
$\rho_{\text{Si$_3$N$_4$}}=3170 kg/m^3$ and specific heat capacities
$c_p^{\text{SiC}}=670 J/kg\cdot K$, $c_p^{\text{SiO$_2$}}=2650 J/kg
\cdot K$, $c_p^{\text{Si$_3$N$_4$}}=673 J/kg\cdot
K$~\cite{adachi2004handbook}, and taking into account volume
fractions, we obtain $m_t=2.22\times 10^{-10}J/K$. Assuming four
insulating posts of length $1\mu m$ and cross section $0.1\mu m \times
0.1\mu m$, each made of SiO$_2$ of thermal conductivity
$\kappa=1.4W/m\cdot K$, thermal conductance $G=5.6\times 10^{-8}W/K$
is calculated based on Fourier's law of conduction. While we
considered insulating posts for simplicity of this calculation, the
bridge structure typically used in standard thermal IR detectors can
also be employed without any changes to the detector pixel design. In
fact, advances in microelectromechanical systems (MEMS) technology
have now enabled $G \lesssim 10^{-8} W/K$ using such bridge structures
for thermal IR detectors~\cite{kimata2018uncooled}. The response speed
is characterized by the time constant $\tau=m_t/G$ which is $4ms$ for
our design. For typical microbolometers, the absorbed thermal power
depends on the pixel area $A_p$, pixel absorptivity plotted in
Fig.~\ref{fig1}(d), the transmissivity of the lens used for focusing
the IR light, temperature and distance of the object emitting the IR
light etcetera. Instead of introducing many new parameters here, we
refer to the previous work and use the typical values of $P_t$ for
typical microbolometers of similar pixel areas which are in the range
of tens of nW~\cite{svatovs2018precise,ryu2015thermal}.

\begin{figure}[t!]
  \centering\includegraphics[width=\linewidth]{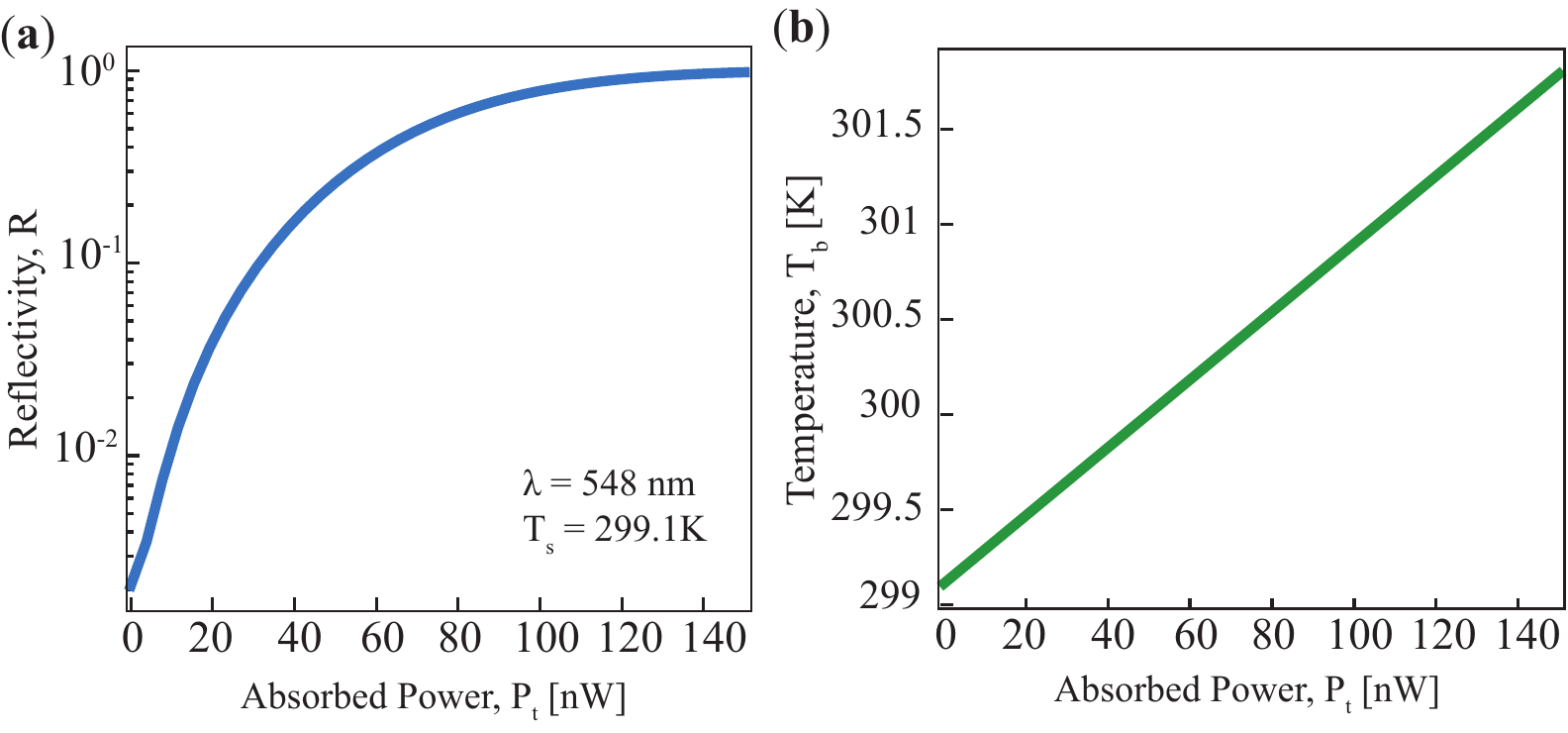}
  \caption{(a) For a fixed substrate temperature $T_s$ and
    illumination wavelength, the reflectivity $R$ increases
    monotonically with increasing absorbed thermal IR power. The
    underlying increase in the detector pixel temperature $T_b$ is
    shown in (b). These changes are obtained by solving Eq.\ref{heat}
    in the main text.}
  \label{fig2}
\end{figure}

Figure~\ref{fig2}(a) depicts the change in the reflectivity for the
specific example system analyzed above as a function of absorbed
thermal IR power. We assume that the intensity of incident visible
light coming from the diffractive optical waveguide is around $2W/m^2$
(approximately $800Cd/m^2$ similar to the brightness levels of
displays like computer monitors). Because of small absorptivity $A
\sim 0.01$ of the pixels of area $A_p=144 \mu m^2$, the absorbed
optical power is $3pW$ which is much smaller than the typical values
of absorbed thermal IR power. The heating due to thermal IR absorption
leads to small changes in the detector temperature ($T_b$) which is
demonstrated in Fig.~\ref{fig2}(b). Accordingly, the reflectivity is
changed leading to strong reflection corresponding to LWIR radiation
coming from hot objects in the far field. We note that the
reflectivity changes over wider temperature ranges of $T_b$ can also
be engineered with different optimized nanophotonic design. The
possible temperature swing of the detector pixel primarily depends on
the thermal conductance $G$ and the maximum absorbed thermal IR power
by the relation $\Delta T_b = \text{max}\{P_t\}/G$. These parameters
can be estimated in separate experiments and the design can be
iterated to realize optimized performance.

\begin{figure}[t!]
  \centering\includegraphics[width=\linewidth]{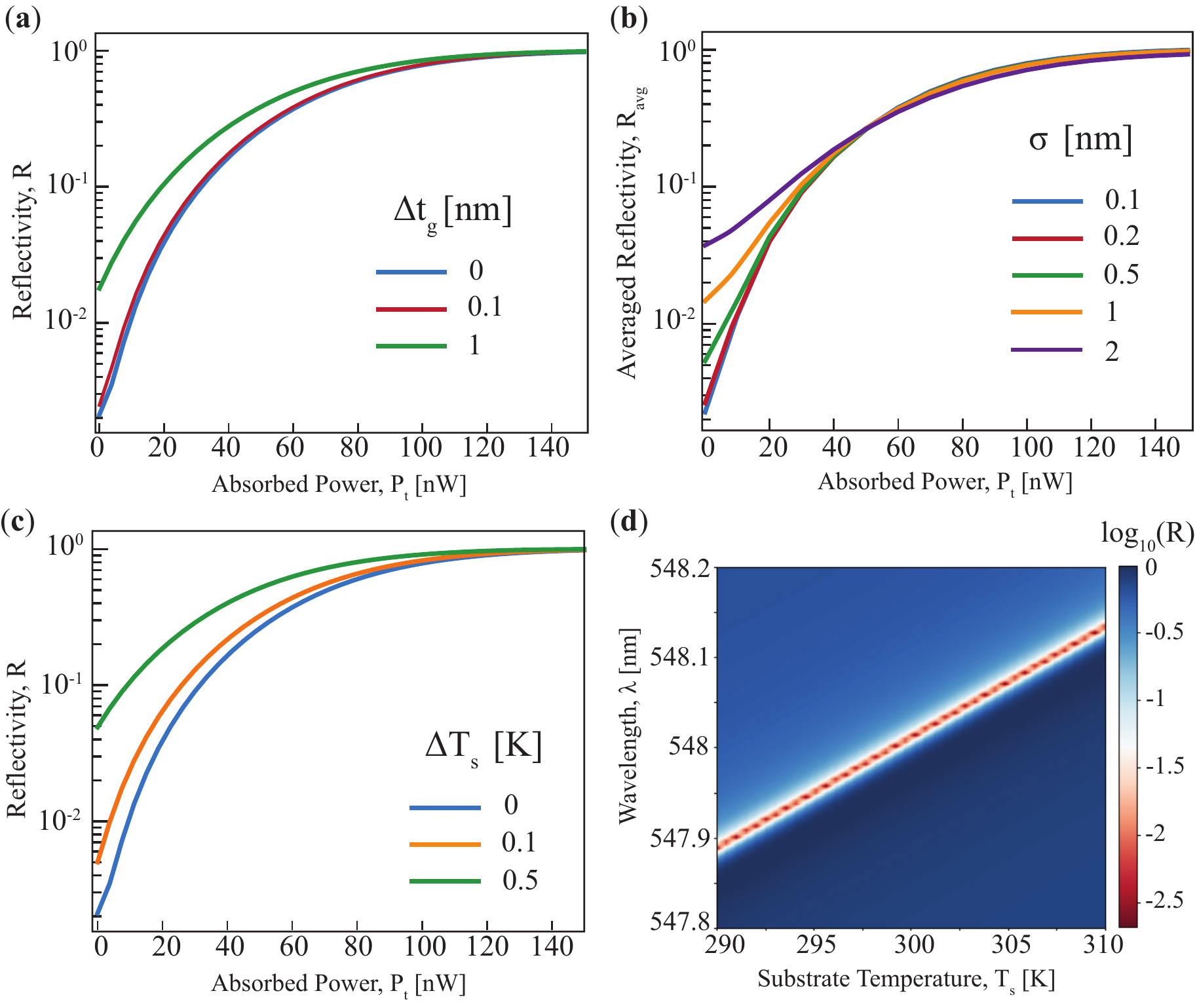}
  \caption{(a) Reflectivity variation is plotted by introducing
    perturbation in the grating height ($\Delta t_g$). (b) Since the
    fabricated design will have nonuniformities related to etching and
    patterning, we assume a random Gaussian perturbation of zero mean
    and standard deviation $\sigma$ in the grating height of each
    detector pixel. The averaged reflectivity plotted in the figure
    for $\sigma=2 nm$ indicates the monotonic increase of $R$ for all
    pixels which is sufficient for producing a reliable IR image using
    the detector array. (c) Reflectivity variation is plotted by
    introducing variation in the substrate temperature denoted as
    $\Delta T_s$. Large deviation for small $\Delta T_s=0.5K$
    indicates the requirement of either $T_s$ stabilization or
    wavelength calibration. (d) Contour plot of reflectivity as
    function of wavelength and $T_s$ reveals the resonant dips (red)
    corresponding to the condition $R \approx 0$. By calibrating the
    operating wavelength to approximately follow this condition based
    on real-time measurement of $T_s$, significant variations due to
    temperature drifts can be avoided.}
  \label{fig3}
\end{figure}

We note that the pixel uniformity across all pixels is important for
recreating reliable final IR image of the far-field scene. The
proposed design which uses high $Q$ resonance for refractive index
sensing, also responds sensitively to design perturbations. For
example, Fig.~\ref{fig3}(a) demonstrates the reflectivity dependence
on $P_t$ by introducing small perturbation $\Delta t_g$ in the
thickness of the grating. As evident from the figure, nanometric
fabrication variations can cause reflectivity to deviate for small
values of $R$. While lattice periods are usually accurate, the grating
structures may experience small perturbations even for a single
pixel. To address these perturbations, we assume a zero-mean
Gaussian-distributed error of standard deviation $\sigma$ in the
grating thickness for all detector pixels and plot the averaged
reflectivity in Fig.~\ref{fig3}(b). At small values of $P_t$, the
averaged reflectivity at $\sigma \approx 2nm$ deviates from the
averaged reflectivity for the design with small values of $\sigma
\approx 0.1nm$. Although such deviations need to be analyzed for
accurate quantitative estimations of temperatures of far-field
objects, from the perspective of direct thermal IR vision which is the
focus of this work, high contrast and monotonic increase exhibited by
the averaged reflectivity of the detector pixels having nanometric
fabrication nonuniformity can be considered acceptable. For the
proposed design, the detector FPA of $1024 \times 1024$ pixels will
require fabrication precision of at most few $nm$ standard deviation
of surface roughness over $\sim mm^2$ area of the device. We note that
advances in lithography techniques such as nanoimprint lithography
have enabled low-cost fabrication of large-area patterned
surfaces~\cite{sreenivasan2017nanoimprint}. In a separate
work~\cite{fan2020high}, subnanometer precision in terms of surface
roughness of approximately $0.15nm$ has been realized for SiC on
insulator surfaces. Therefore, we believe that fabrication
requirements for the proposed design can be ensured using
state-of-the-art lithography techniques.

In the design of microbolometer IR detectors, significant importance
is given to drifts or small changes in the substrate temperature
($T_s$). The detector is designed to be highly sensitive to $T_b$
which depends on the absorbed IR power and the substrate temperature
$T_s$. Variation in $T_s$ causes an equal variation in $T_b$, and this
becomes a major source of unwanted signal which is unrelated to IR
signals coming from the far field. Various calibration or compensation
approaches are used in standard microbolometer IR detectors to tackle
this important issue~\cite{lv2014uncooled,nugent2013correcting}. The
same holds true for the proposed design. For instance,
Fig.~\ref{fig3}(c) demonstrates the reflectivity dependence on $P_t$
for small variations in the substrate temperature ($\Delta T_s$) for
fixed illumination wavelength. Evidently, the deviation in $R$ is
quite large for small $\Delta T_s \sim 0.5K$.  This problem can be
solved by tuning the illumination wavelength in real time based on
prior measurements of reflectivity for different values of $T_s$. The
contour plot in Fig.~\ref{fig3}(d) illustrates that the illumination
wavelength which should coincide with the resonant dip ($R \lesssim
10^{-3}$) in the reflectivity spectrum increases linearly with
$T_s$. Such linear tunability can be achieved with on-chip
microheaters with integrated tunable visible laser
sources~\cite{franken2020hybrid,boller2020hybrid,malik2021low}. While
our proposed architecture does not involve electronic readout and
digital processing, it will require a thermocouple to measure $T_s$
and a controller system to tune the illumination wavelength in a
predetermined manner in real time. An alternative to such a controller
system can be temperature stabilization using small Peltier elements
used in some versions of thermal IR cameras. On-chip device
temperatures can be stabilized within few $mK$ temperature
variations~\cite{choi2003design} and the operating wavelength can be
fixed to coincide with the resonant dip in the reflectivity
spectrum. We note that the fabrication nonuniformity and the
temperature drifts can also be mitigated by using illumination light
of bandwidth comparable or close to the bandwidth of the resonant dip
in the reflection spectrum instead of using monochromatic light. Since
the response is integrated (or averaged) over a bandwidth of
wavelengths, the sensitivity to unwanted design perturbations is
reduced. Aside from the specific detector pixel requirements and
considerations related to fabrication and temperature drifts
(discussed above), vacuum packaging and radiation shields are required
for developing a proof-of-concept thermal IR imaging system. The same
packaging solutions used for traditional microbolometer
FPAs~\cite{garcia2017radiometric} can be translated and used for our
design.

\section{Conclusion}

We introduced a new nanophotonic detector in place of traditional
microbolometer as a viable minimalist approach to thermal IR
imaging. Our proposed design produces an IR image as spatially
modulated wavefront of reflected visible light which can be directly
viewed by the human eye. This design eliminates the need for
additional components for electronic readout, A/D conversion, image
processing and information display, otherwise necessary in standard
thermal IR cameras. The proposed design can be useful for achieving
the same functionality as IR cameras potentially at reduced cost, form
factor and power consumption. We note that digital processing of
electronic or optical measurement of the detector pixels is useful for
additional error-compensation techniques and is necessary for
providing quantitative estimates (in numbers) of temperatures of
far-field objects. A figure of merit called as noise equivalent
temperature difference (NETD) is used to compare the performance of IR
cameras in their ability to quantify tiny temperature differences
between two far-field
objects~\cite{gade2014thermal,rogalski2017next}. Our design does not
provide such quantitative temperature estimates since it is designed
to directly provide an IR image to the human eye. Nonetheless, there
can be use cases where the proposed minimalist approach can be
potentially employed. For many practical applications of IR cameras
such as inspection and detection in industrial, infrastructure,
agriculture, healthcare settings, night vision or low-light vision for
military, fire-fighters, camping enthusiasts, the objective of the
device is to provide an IR image to the human eye. In some of these
use cases, the final human evaluation is often based on the IR image
and not the temperature numbers, thus requiring strong contrast and
good quality of the IR image. Therefore, we believe that the proposed
minimalist approach can be practically useful. 

Apart from the actual proof-of-concept experimental demonstration,
there are many interesting aspects of this work that can be further
extended in the near future. All components used in the design are
transparent at visible wavelengths. If the optical lenses are
optimized simultaneously for visible and infrared wavelengths to
reduce chromatic aberration, it is conceivable that the final image
will contain IR image at the specific laser illumination wavelength
overlaid on top of the visible image. Such augmented thermal vision
functionality with frugal resources can be an interesting new
technology. One can also design multi-spectral IR imaging where
different IR wavelengths are mapped to different visible wavelengths
using multiple optimized pixels. For example, two types of detector
pixels can be designed to potentially map SWIR ($3-8\mu m$) light to
green light and LWIR ($8-14\mu m$) light to red light. Such a
functionality is quite challenging to develop even with traditional
microbolometer detectors. We believe that our work paves the way for
developing such novel IR imaging capabilities in the near future. \\
 
\section{Methods}

The permittivities of SiC, SiO$_2$ and Si$_3$N$_4$ are obtained from
various references~\cite{palik1998handbook,luke2015broadband}. The
simulation of reflection and transmission is performed using
well-known rigorous coupled wave analysis (RCWA)
technique~\cite{moharam1995formulation,jin2020inverse}. The
open-source code is available at
\href{https://pypi.org/project/grcwa/}{GRCWA} under GPL license. For
the calculation of the detector pixel absorptivity of the incident IR
radiation plotted in Fig.\ref{fig1}(d), we calculate the flux
reflected from the pixel and the flux in the vacuum gap between the
pixel and the substrate, assuming normal incidence of light. For our
nanophotonic design of the pixel as a refractive index sensor, we use
the Nelder-Mead method to find the geometric parameters (thickness of
each material layer) which minimize the ratio of the reflectivities
given by $R(T_b=300K)/R(T_b=305K)$ at a fixed wavelength. Because of
physics-informed ordering of layers (intended to support guided modes
inside SiC and Si$_3$N$_4$ layers), this optimization upon few trials
and starting points yields optimized design with $R(T_b=300K) \lesssim
10^{-3}$ and $R(T_b=305K) \sim 0.9$. Additional inverse design
constrained and informed by nanofabrication capabilities and costs can
certainly be formulated to automate this design with the objective of
achieving the reflectivity change from $R\approx 0$ to $R \approx 1$
over a specific temperature differential. This differential is given
by $\Delta T_b=\text{max}\{P_t\}/G$ where the conductance $G$ and the
typical absorbed thermal IR power $P_t$ for the detector pixel can be
estimated separately for the fabricated design.

\section{Acknowledgement}

C.K. would like to acknowledge helpful discussion with Dr. Bo Zhao
regarding the potential use of hexagonal boron nitride (hBN) in our
nanophotonic design. This work is supported by a MURI project from the
U.S. Air Force Office of Scientific Research (FA9550-21-1-0312). 

\bibliography{photon}

\end{document}